\documentclass{amsproc}

\theoremstyle{definition}

\theoremstyle{remark}

\numberwithin{equation}{section}

\newcommand{\be}{\begin{equation}}
\newcommand{\ee}{\end{equation}}
\newcommand{\hf}{{1\over 2}}
\newcommand{\tr}{\it tr}
\newcommand{\p}{\partial}
\newcommand{\la}{\lambda}

\begin{document}

\title{Vortex condensation and black hole  \\ 
in the matrix model of 2-d
       string theory }

\author{Vladimir A. Kazakov}
\curraddr{Laboratoire de Physique Th\'eorique de l'Ecole Normale Sup\'erieure,
24 rue Lhomond, 75231 Paris Cedex05, France}
\email{kazakov@physique.ens.fr}

\date{April 30, 2001}


\begin{abstract}
We review recent results in the matrix model approach to the 2-d
noncritical string theory compactified in time, in the phase of
condensation of the world sheet vortices (above the
Berezinski-Kosterlitz-Thouless phase transition)
\cite{KKK,KATS,ALKA}. This phase is known to describe strings on the
2-d black hole background, due to the conjecture of V.Fateev, A. and
Al.Zamolodchikov. The corresponding matrix model has an integrable
Toda structure which allows to compute many interesting physical
quantities, such as string partition functions of various genera and
the 1- and 2-point correlators of vorticities of arbitrary charges.
(Talk presented at Strings 2001, Mumbai, India)

\end{abstract}

\maketitle

\section{Introduction}

String theory formulated as a CFT on the world sheet can be
perturbed, as any CFT, by marginal operators.  The RG flows in the
target space corresponding to these perturbations can bring the theory
from the initial flat target space to the fixed points corresponding
to curved backgrounds, such as black holes. Whereas in critical
strings the task of describing such a flow is usually very difficult
it can be feasible in the context of a ``toy'' noncritical string theory
in two target space dimensions.
 
An example of such  nontrivial flow was given in \cite{FZZ} who
considered the so called Sine-Liouville (SL) theory with the
Lagrangian
\begin{equation}
\label{SL}
L={1\over4\pi}\left[(\partial x)^2 +(\partial\phi)^2
-2Q\hat R\phi+\lambda e^{b\phi}\cos R(x_L-x_R)\right].
\end{equation}
The central charge of this theory is $c=2+6Q^2$, the compactification
radius of the bosonic field $x=x_L+x_R$ fixed as $R^2=2+Q^{-2}$ and
$b=-1/Q$ for the Sine-Liouville perturbation to be marginal. The
perturbation here is just the sum of two lowest winding modes (with
winding charges $\pm 1$). For $\lambda=0$ this is a linear sigma model
describing the simplest compactified version of the non-critical 2-d
string theory.
\footnote{Strictly speaking the Lagrangian \ref{SL} describes the
usual non-critical $c_M=1$ 2-d string theory only for $R=3/2$ when
$c=26$, unless one changes the matter field content.}  When one turns
on $\lambda$ the Sine-Liouville perturbation drives the theory to a
new IR fixed point in the $(x,\phi)$ target space characterized by a
high density of windings. As was conjectured by \cite{FZZ} this fixed
point corresponds precisely to the level $k$ WZNW CFT on the coset
${SL(2,C)\over SU(2)\times U(1)}$ with the central charge $c={3k\over
k-2}-1$. For $k=9/4$ we have $c=26$ and this sigma model describes the 2-d
string theory in  dilatonic Euclidean black hole (``cigar'') background.

Already superficially, both theories have many similar features: their
asymptotic background at $\phi\to\infty$ is just a cylinder of the
radius $R=\sqrt{k}$ in $(X,\phi)$ coordinates, so the asymptotic
states are the same. In the Sine-Liouville theory they are defined by
the vertex operators:
\begin{equation}
\label{OPER}
V_{j,n_1,n_2}=\exp [ip_Lx_L+ip_Rx_R+2Qj\phi] 
\end{equation}
with $P_{L,R}=R^{-1}n_1+R n_2$, where $n_1$ and $n_2$ are arbitrary
integers. These operators have the conformal dimensions:
\begin{equation}
\Delta_{j,n_1,n_2}=-{j(j+1)\over R^2-2}+{1\over
4}\left(R^{-1}n_1+R n_2\right)^2
\label{DIM}
\end{equation}
These dimensions coincide with the dimensions of the corresponding
operators on the coset model side. 

It was also noticed by \cite{FZZ} that both theories have coinciding
2- and 3-point correlators of these operators (some of them were
already calculated in \cite{FAT} and \cite{TESCH}).

Qualitatively, the reason for the same physics in these two theories
lies in the fact that both have a ``wall'' in the target space which
prevents the propagation of the states beyond a certain point along
the axis of the cylinder: in the cigar model this wall corresponds
just to the tip of the cigar, whereas in the theory (\ref{SL}) it is
the SL potential which plays the role of the ``wall''. The theory is
strongly coupled in the region $\phi<0$, and although the SL potential
is penetrable classically (due to the $\cos$ oscillations) it appears
to be unpenetrable due to the quantum effects.

The conjectured equivalence is an example of the strong-weak duality:
at large $k$ the curvature of the cigar becomes small, thus the theory
is weakly coupled, whereas the SL wall becomes steeper and the
interaction is concentrated in the strong coupling region near this
wall. 

A strong piece of evidence of this duality was brought in the recent
paper \cite{HORI} where the equivalence was proven for the $N=2$
supersymmetric analogues of these models and appeared to be an example
of the mirror symmetry.

We conclude that although the straightforward proof of the conjecture
of \cite{FZZ} is still missing in the bosonic case we have enough of
evidence to view the SL theory (\ref{SL}) as an alternative model for
the description of the dilatonic black hole (at least at $R=3/2$).

On the other hand, the usual Liouville theory at the central charge
$c=26$ is known to be very effectively described and studied in the
matrix quantum mechanics (MQM) formulation of the noncritical 2-d
($c_M=1$) string theory. The SL perturbation amounts to introducing
the vortices of the charge $\pm 1$ on the world sheet of the string
(similar to the vortices in the usual Sine-Gordon model). Since we
know how to implement vortices on the MQM language
\cite{GRKL1,GRKL2,BULKA} we can  formulate and study the strings
in the 2-d black hole background by means of powerful matrix methods.

\section{ Black hole as a perturbation of $c_M=1$ string }

Let us turn to the noncritical 2-d bosonic string perturbed by the SL
term, described by the Lagrangian
\begin{equation}
\label{STR}
L={1\over4\pi}\left[(\partial x)^2 +(\partial\phi)^2
-4\hat R\phi+\mu\phi e^{-2\phi}+\lambda e^{(R-2)\phi}
\cos R(x_L-x_R)\right]
\end{equation}
Here the paprameters are chosen in such a way that the total central
charge is always $c_{tot}=26$ and the SL term and the cosmological
term are marginal to enforce the conformal invariance, whereas the
compactification radius $R$ can be arbitrary. Note that at $R=3/2$ and
zero cosmological constant $\mu=0$ we have again the SL theory
(\ref{SL}) at the black hole point $Q=2$. Hence the theory (\ref{STR})
is just a different (comparing to the WZNW or the SL models)
deformation from the conventional 2-d black hole with respect to the
parameters $\mu$ and $R$.

From the dependence of the Lagrangian (\ref{STR}) on the zero mode of
the Liouville field $\phi$ we can immediately read off the target
space scaling of physical quantities, such as the SL coupling
$\lambda$ (the fugacity of vortices) and the string coupling $g_s$,
with $\mu$:
\begin{equation}
\lambda\sim \mu^{2-R\over 2}, \ \ \ \  g_s\sim 1/\mu.
\label{SCAL}
\end{equation}

From the matrix model results \cite{KLREW} 
we know that the free energy for $\lambda=0$ has the following
expansion in genera:
\begin{equation}
\label{FRNOT}
\begin{split}
 F(\mu,\lambda=0)& = \log Z_{ \mu}[0] \\
&= {1\over 2\pi} \int _{-\infty}^{\infty} dE\sum_{k=0}^{\infty} {k+\hf
\over E^{2}+(k+\hf)^{2}} \log(1+e^{-2\pi R(\mu-E)})\\
&  = -  {R\over2}\mu^2 \log \mu -
{1\over24}\big(R + {1\over R} \big)\log \mu +
R\sum_{h=2}^\infty \mu^{-2(h-1)} g_h(R)+O(e^{-2\pi\mu}),\\
\end{split}
\end{equation}
where we introduced the polynomials in ${1\over R}$
\begin{equation}
\label{BERN}
 g_h(R)=
(2h-3)! \  2^{-2h} \sum_{n=0}^{h} 
\left({1\over R }\right)^{2n}
{(2^{2(h-n)}-2)(2^{2n}-2)\vert B_{2(h-n)} B_{2n}\vert \over
[2(h-n)]![2n]!}
\end{equation}
and $B_m$ are Bernoulli numbers.  The partition function (\ref{FRNOT})
has a T-duality symmetry, $R \to {1\over R }$, $\mu \to R \mu$.  This
symmetry will be broken in the presence of vortices (i.e. for
$\la\ne 0$).

The first two terms of $1/\mu$ expansion in (\ref{FRNOT})
corresponding to the sphere and torus partition functions of the 2-d
string theory are known as well  from the direct calculations in the
conformal $\sigma$-model (\ref{STR}) (see, for example, \cite{KLREW}).

Using this result and the scaling (\ref{SCAL}) we can represent the
genus expansion for the free energy (``string partition function'')
$F(\lambda,\mu)=\sum_{h=0}^\infty F_h(\lambda,\mu)$ in 2-d string
theory already for $\lambda\ne 0$ in the form:
\begin{equation}
\label{SCEXP}
\begin{split}
F_0(\lambda,\mu)&=-{R\over 2}\mu^2\log\mu+ \mu^2 A_0(z),\\ 
F_1(\lambda,\mu)&=-{1\over24}(R+{1\over
R})\log\mu+A_1(z),
 \ \ \ F_h(\lambda,\mu)=\mu^{2-2h}A_h(z),
\end{split}
\end{equation}
where $z$ is defined as $z=\sqrt{R-1}\lambda\mu^{R-2\over 2}$ and is,
according to the scaling (\ref{SCAL}), a dimensionless parameter.
It is clear that $A_h(0)=f_h(R)$ from (\ref{FRNOT}).

The black hole limit of (\ref{SCEXP}) corresponds to $z\to\infty$.  In
this limit it is rather the SL term in (\ref{STR}) and not the
cosmological term which governs the size of the worldsheet. It means
that the gas of vortices generated by the SL interaction becomes
dense. So the creation of the black hole is closely related to the
condensation of vortices. It is known \cite{GRKL1,GRKL2,BULKA} that
vortices start condensing for $R<2$, i.e. for the temperatures above
the Berezinski-Kosterlitz-Thouless (BKT) phase transition
($R_{BKT}=2$). The 2-d black hole point ($R=3/2$) lies already in the
phase of condensing vortices.

In order to be able to study the black hole physics we have to find
the values of coefficients $A_h(z)$ in (\ref{SCEXP}), or at least
their limiting values $A_h(\infty)$. For that we will use the matrix
model techniques and the integrability properties of the corresponding
MQM.

\section{Matrix quantum mechanics and vortices on the worldsheet}

To describe the gas of vortices on the worldsheets of the 2-d string
we will use the MQM defined by the partition function
\begin{equation}
\label{MQM}
 Z_N(\Omega)
= \int_{M(2\pi R)=\Omega^{\dagger} M(0)\Omega} {\it D}^{N^2}M(x) e^{ - \tr
\int ^{2\pi R} _0 dx \left[ \hf (\p_x M)^2  +V(M)\right] }
\end{equation}
where $M(x)$ is a Hermitian $N\times N$ matrix field depending on the
compact ``time'' $x$, $\Omega\in SU(N)$ is a twist matrix and the
matrix potential can be chosen, for example, as a cubic polynomial:
$V(M)=\hf M^2-{g\over 3 \sqrt{N}}$. Due to the periodic boundary
condition the theory is compactified on the time circle of the length
$2\pi R$, as the coordinate $x=x_L+x_R$ of the model (\ref{STR}).

The planar Feynman graphs generated by the MQM can be viewed as
discretized world sheets of the $c=1$ noncritical string theory
\cite{KAZMIG}. The time $x$ represents one of the two target space
coordinates of the theory (another one - the Liouville coordinate -
being hidden into the matrix structure of the theory (see
\cite{POLCH,JEV}). In the compactified version of the model the vortex
excitations can appear on the discretized worldsheets, in analogy with
the model of planar rotators on a regular lattice used in statistical
mechanics to describe the BKT phenomenon.
The propagators (corresponding to the double index lines) look as follows
\begin{equation}
\label{PROP}
x_v\
^{^i}_{_k}\!\!\!\!\!\!\!\ = \! = \! = \! =\!\!\!\! \!_{_{m}} \!
\!\!\!  = \! = \!  = \! = \!\!\!\! ^{^ j}_{_l}\ x_{v'}=  \sum_{m=
-\infty}^\infty
\exp\left(- |x_v-x_{v'} +2\pi R m |\right)(\Omega^m)_{ij} (\Omega^{-m})^{kl},
\end{equation}
where $x_v$ is a coordinate of the graph vertex $v$ on the $x$-circle
and $m$ is the number of windings of the propagator around
the circle (in a given term in the sum).

Applying the Feynman rules for the double-lined graphs we obtain the
following formal expression for the free energy $F_N(\Omega)=\log
Z_N(\Omega)$ as a sum over graphs $G_n^{(h)}$ and their genera $h$ and
sizes $n$ (the number of cubic vertices in the graph):
\begin{equation}
\label{SUMH}
\begin{split} 
&F_N(\Omega) =\\
&\sum_{h=0}^\infty N^{2-2h}\sum_{n=2}^\infty g^{n}
\sum_{G_n^{(h)}}\ \prod_{f\in G}    
{\tr \Omega ^{w_{f}}\over N} \sum_{\{m_{vv'}\}}\int\prod_{v\in G} d\ x_v
\prod_{<\!vv'\!>}\exp-\left(|x_v-x_{v'} +2\pi R m_{vv'}| \right),
\end{split}
\end{equation}
where $w_f= \sum_{\ell\in \p f} m_{\ell}$ is the vorticity flowing
through the face $f$ of a graph. 

The partition sum on any fixed graph $G_n^{(h)}$ (including the sum
over $m$'s and integrals over $x_v$'s with the last exponential
factor) resembles very much the Villain model \cite{VILL}. Originally
the Villain model was formulated on a regular square lattice and one
had a square of the expression in the exponent. However, the most
important feature to retain here is the periodicity in $x_v$
variables, so both models should be in the same class of universality
(different of course for regular and random graphs).  We conclude that
the MQM (\ref{MQM}) describes the gas of vortices living on faces of
dynamical lattices. The weights ${\tr \Omega ^{w_{f}}\over N}$ play
the role of fugacities of vortices with different charges $w_{f}$.

The continuum limit of the model corresponds, as usually, to tuning
$g\to g_c$ where the characteristic graphs become very big and play
the role of continuous world sheets of the Polyakov string. The sum
over graphs simulates the sum over 2-d metrics, and the integrals over
$x_v$ represent the functional integral over the bosonic field.

In the CFT language, we can represent the analogous model (at least in
the weak coupling regime) by the Lagrangian
\begin{equation}
\label{STRT}
 L={1\over4\pi}[(\partial x)^2 +(\partial\phi)^2
 -4\hat R\phi+ \mu\phi e^{-2\phi}+
 \sum_{n\ne 0}  t_{n}  e^{(  |n|R-2)\phi}  
e^{i n R(x_L-x_R) }].
\end{equation}
where $t_n\sim {\tr \Omega ^n\over N}$ up to some subtleties discussed
below. The Lagrangian (\ref{STRT}) generalizes (\ref{STR}) to the
perturbations by vortices of arbitrary charge.  However, as we see
from the Liouville field dependence of the $n$'th term in
(\ref{STRT}), only the vortices of charge $\pm 1$ are relevant in the
target space in the interval $1<R<2$ which we will be interested in.

\section{ Integration over the twist angles }

As we will see later, it is advantageous instead of fixing the twist
variable  $\Omega$ to integrate over it with a special weight:
\begin{equation}
\label{INOM}
Z_N[ \lambda]= \int [D\Omega]_{SU(N)} \exp\left( \sum_{n\in {\bf Z}} \lambda
_n \tr \Omega^{n}\right)  Z_N(\Omega).
\end{equation}
In this way we change the original matrix variable $\Omega$ to the
infinite set of new variables $\lambda_n, \ \ n\in {\bf Z}$.  We see
from (\ref{INOM}) that under the $U(1)$ shift $\Omega\to
e^{i\alpha}\Omega$ the couplings $\lambda_n$ transform as
$\lambda_n\to e^{-i\alpha n}\lambda_n$, i.e. in the same way as the
fugacity of a vorticity ${\tr \Omega ^n\over N}$ in a face.  This
suggests that the couplings $\lambda_n$ play the same role in the MQM
formalism as the couplings $t_n$ in the Lagrangian (\ref{STRT}), i.e.
$\lambda_n\sim t_n$ in the continuous limit. A more precise statement
is based on the observation made in \cite{DOUG} who noticed that for
large enough $N$ and $\lambda_n<{\sim} N$'s the integration rules for
the unitary matrices can be written as
\begin{equation}
\label{DOUGL}
\int D\Omega_{SU(N)}e^{\sum_{n\ne 0} \la_n\tr\Omega^n}=e^{\sum_{n\ne 0}
n\la_n\la_{-n}} + O(e^{-N}).
\end{equation}
So the variables $\tr\Omega^n$ look like independent Gaussian
variables in this approximation. 

If we substitute (\ref{SUMH}) into (\ref{INOM}) and start integration
over $\Omega$ according to the rules (\ref{DOUGL}) we will get 3 kinds
of contributions: first, the terms in the exponent in (\ref{INOM})
will couple to each other and give a trivial overall factor equal to
the r.h.s. of (\ref{DOUGL}); second, the $\tr\Omega^n$ will couple to
${\tr \Omega^{w_f}\over N}$ (at $w_f=-n$) which will result in the
substitution $\tr \Omega^{w_f}\to \lambda_{w_f}$, thus giving to
$\lambda_n$ the meaning of the fugacity of vortices of charge $n$; and
third, $\tr \Omega^{w_f}$ in (\ref{SUMH}) will couple to each other.
This last contribution is not yet very well understood (see
\cite{KKKsmall} for the details). On the one hand, one can give some
arguments that it is inessential in the double scaling limit
considered below; on the other hand, it gives some new configurations
of planar graphs (various remote faces can glue together in this way)
which might be important to the right combinatorics of the discretized
``worldsheets''.

We conclude this section by stating that the partition function
(\ref{INOM}) is a good candidate for the matrix model description of
the 2-d string theory with vortices on the world sheet. The couplings
$t_n$ of vortex operators of charge $n$ are proportional to $\la_n$.

\section{ Double scaling limit}

The MQM formulated in the previous section is a complicated theory. It
would be hard to try to perform the integrals over $M(x)$ or over
$\Omega$ directly in (\ref{MQM}). On the other hand, we can
significantly simplify the model by going to the double scaling limit
which means that we are not interested any more in the detailed
structure of the planar graphs but rater want to concentrate ourselves
on very big graphs giving a good approximation to the smooth
worldsheets. In the matrix action it corresponds to the fact
\cite{DSL} that the only relevant part of the potential in (\ref{MQM})
is concentrated around the maximum. If one shifts $M\to M+1/g$ the
quadratic term reverses the sign and the cubic term plays the role of
a steep wall at a distance $\sim \sqrt N$ from the top of the
quadratic maximum. Thus we obtain instead of (\ref{MQM}) a MQM model
of the ``inverted oscillator'', i.e. with the upside-down quadratic
potential:
\begin{equation}
\label{MQMDSL}
 Z_N(\Omega)
= \int_{M(2\pi R)=\Omega^{\dagger} M(0)\Omega} {\it D}^{N^2}M(x) e^{ - \tr
\int ^{2\pi R} _0 dx \left[ \hf (\p_x M)^2  -\hf M^2\right] }.
\end{equation}

We can calculate the functional integral over $M(x)$ by
formally continuing (\ref{MQMDSL}) analytically to the usual stable
harmonic oscillator and than changing $R\to iR$ in the final
answer. One can do the integration by the mode expansion on the circle
separately for each matrix element $M_{kj}(x)$ (they are completely
decoupled before the integration over $\Omega$) with the twisted
boundary condition $M_{kj}(2\pi R)={z_j\over z_k} M_{kj}(0)$ ($k,j$
are fixed), where we took, without a loss of generality, the twist
matrix in the diagonal form $\Omega={\it diag}(z_1,z_2,\cdots,z_N)$.

It is also natural in the double scaling to do the grand canonical
transformation of (\ref{INOM}) passing from fixed $N$ to fixed
conjugated variable - chemical potential $\mu$. The final formula for
the partition function of the model in terms of the integral over
twist angles (taken with the Haar measure for the Cartan subgroup
${1\over N!}\oint \prod_{k=1}^N{d\ z_k\over 2\pi iz_k}
\prod_{m>j}|x_m-x_j|^2\cdots$) looks as follows
\begin{equation}
\label{GRANF}
\begin{split}
&Z[\mu,\lambda]\equiv e^{F(\mu,[\lambda])}=\\
&\sum_{N=1}^\infty{e^{-2\pi R\mu N}\over N!}
\oint \prod_{k=1}^{N} \ {d\ z_k\over 2\pi i z_k} \
{e^{ u(z_k)} \over ( e^{i\pi R}- e^{-i\pi R} ) } \ \prod_{ j \ne j' }^{N} 
{z_j \ -\ z_{j'} \over e^{i\pi R}\ z_j - e^{-i\pi R}\ z_{j'}}.
\end{split}
\end{equation}
where $u(z)=\sum_n \lambda_n z^n$.

The integrals in this formula look very divergent because of the poles
in the last product. Hence we will employ a strategy different from
the direct calculations of these integrals and rather will obtain a set
of differential equations on this partition function.

\section{ The partition function as a $\tau$-function of Toda integrable 
hierarchy} 

It is easy to recognize in (\ref{GRANF}) a particular case of the
$\tau$-function of Toda integrable hierarchy $\tau_l[t]$ defined on
the fermionic vacuum of charge $l$, as it was done in \cite{HKK}. For
that it is sufficient to compare it with the general solitonic
solution of this hierarchy given, for example, in \cite{JIMI}. The
relation looks as follows:
\begin{equation}
\label{TAUF}
Z(\mu-il,[\lambda])= e^{\sum _n nt_{n}t_{-n}}\tau_l[t]
\end{equation}
where $\lambda_n=2i\sin(\pi R n) t_n$. Note that the dependence on $l$
is absorbed into the imaginary part of $\mu$.

The Toda $\tau$-function satisfies a set of Hirota bilinear equations
which can be encoded into the identity
\begin{equation}
\label{HIROTA}
\begin{split}
&\oint_{C_\infty}{dz\over 2\pi i} z^{l-l'}
\exp\Big(\sum_{n>0} (t_n-t'_n) z^n\Big)
\tau_{l}(t-\tilde\zeta_+)
\ \tau_{l'}(t' +\tilde\zeta_+ ) =\\
& \oint_{C_0}{dz\over 2\pi i} z^{l-l'}
\exp\Big(\sum_{n<0} (t_n-t'_n) z^{-n}\Big) \tau_{l+1}(t - \tilde\zeta_-)
\ \tau_{l'-1}(t' +\tilde\zeta_-) .
\end{split}
\end{equation}
where
$\tilde \zeta_+ = (\ldots, 0, 0, z^{-1}, z^{-2}/2, z^{-3}/3 , \ldots),
\ \ \tilde \zeta_- = (\ldots, z^{3}/3, z^{2}/2, z , 0, 0  \ldots).
$
Expanding in $y_{n} = t'_{n} - t_{n} $ we obtain an infinite hierarchy
of soliton partial differential equations.  The lowest, Toda equation,
which is the most important for us, is obtained as
the coefficient in front of $y_{-1}$:
$$\tau_l\p_1\p_{-1}\tau_l-\p_1\tau_l\p_{-1}\tau_l+
\tau_{l+1}\tau_{l-1} =0.$$ Written directly in physical variables
$\mu,\lambda_{\pm 1}$ for the free energy $F(\mu,[\la])$ it looks as
\begin{equation}
\label{TODA}
{\p\over\p \lambda_{+1}}{\p\over\p\lambda_{-1}} F(\lambda,\mu) +\exp
 \left[ F(\lambda,\mu+i) + F(\lambda,\mu-i)-2 F(\lambda,\mu) \right]
 =1.
\end{equation}
This equation describes the dynamics of vortices of the lowest charges
$n=\pm 1$ and hence is appropriate for studying the 2-d string
 with the Sine-Liouville type interaction (\ref{STR}).  We
will put for the moment $\lambda_n=0, \ \ n=\pm 2,\pm 3,\cdots$ and
denote $\lambda_{\pm 1}=e^{\pm i\alpha}\lambda$. Note also that due to the
vortex charge neutrality of the system the free energy does not depend
on the phase $\alpha$ and is only a function of two variables $\mu$
and $\lambda$ (and a parameter $R$) which we denote just as
$F(\mu,\lambda)$. It satisfies the eq. 
\begin{equation}
\label{TODAR}
{1\over 4}\lambda^{-1}\p_\lambda\lambda\p_\lambda F(\mu,\lambda) +\exp
 \left[ F(\mu+i,\lambda) + F(\mu-i,\lambda)-2 F(\mu,\lambda) \right]
 =1.
\end{equation}

This equation should be supplied by an appropriate boundary
condition. It is known \cite{KLREW} that in the double scaling limit
$F(\mu,\lambda=0)$ as defined in (\ref{GRANF}) is given by the
Gross-Klebanov formula (\ref{FRNOT}) with the cosmological coupling
$\mu$ expressed in the matrix model through the chemical potential of
the matrix eigenvalues which appear to be free fermion coordinates in
the upside down matrix potential in this case.  So we can take
(\ref{FRNOT}) as a boundary condition. Another necessary boundary
condition stems from the fact that $F(\mu,\la)$ has a regular
$\lambda^2$ expansion at $\lambda=0$ since it is just an expansion in
the number of vortex-antivortex pairs.

Let us note here that the eq. (\ref{TODAR}) has zero modes for any
coefficients $C_n,D_n$:
$$
\Delta F=   \sum_{n=0}^\infty
(C_n+D_n  \log \lambda  )\  e^{ -2\pi\mu n}.
$$
It means that nonperturbatively this equation does not fix the whole
solution. The similar $O(e^{ -2\pi\mu})$ terms are present in the
boundary condition (\ref{FRNOT}), together with the terms $O(e^{
-2\pi\mu R})$ following from the T-duality of (\ref{FRNOT}).

\section{Partition functions for fixed genera}

We are ready now to calculate $F(\mu,\lambda)$, at least
perturbatively in powers of $g_s\sim 1/\mu$. The finite
difference operator in (\ref{TODAR}) will be understood as an
expansion in powers of $\p_\mu$ as well. For example, in the spherical
limit the eq.(\ref{TODAR})  becomes
\begin{equation}
\label{TODAS}
{1\over 4}\lambda^{-1}\p_\lambda\lambda\p_\lambda F(\mu,\lambda) +\exp
 \left[ -\p_\mu^2 F(\mu,\lambda) \right]
 =1.
\end{equation}

Plugging the expansion (\ref{SCEXP}) into (\ref{TODAR}) and analyzing
it order by order in $1/\mu$ we obtain the following results for the
individual genera:

{\bf (i) Genus 0}
\begin{equation}
\label{FZERO}
\p_\mu^2 F_0(\mu,\lambda)=-{2R\over 2-R}(\log(\la\sqrt{R-1})
+ X_0(y)
\end{equation}
where $X_0(y)$ satisfies the equation
\begin{equation}
\label{SUSC}
y= e^{-{1\over R} X_0} - e^{ {1-R\over R} X_0}
\end{equation}
where we used the variable 
$y=z^{2\over R-2}=\mu\left(\sqrt{R-1}\lambda\right)^{-2\over 2-R}$.

Using this solution we can expand $F_0(\mu,\lambda)$ in powers of
$\lambda^2\mu^{R-2}$ corresponding to the contributions (correlators)
of the vortex-antivortex pairs reproducing the formula
conjectured long ago by G.Moore \cite{MOORE}:
\begin{equation}
\label{MOORE}
F_0(\lambda,\mu)=-{R\over2}\mu^2\log\mu+R\mu^2
\sum_{n=1}^\infty{1\over n!}\left[\mu^{R-2}\tilde\lambda^2\right]^n
{\Gamma(n(2-R)-2)\over \ \Gamma(n(1-R)+1)}.
\end{equation}
where $\tilde\lambda=\sqrt{R-1}\la$.  It is remarkable that in the
interval
\footnote{ Note that for $R<1$ the singularity defining the
convergence radius of this series in terms of $z$ variable becomes
real; as was explained in \cite{HSUK} this critical point corresponds
to the trivial $c_M=0$ in the IR regime. However it is not the case
for the $1<R<2$ interval where we will encounter a new ``black hole''
phase.}  $1<R<2$ we can also employ another expansion - in powers of
${\mu\over\lambda^{2\over 2-R}}$, thus encountering the black hole
limit of our model:
\begin{equation}
\label{BHEXP}
\begin{split}
&F_0(\lambda,\mu)=-{(2-R)^2\over R-1}\tilde\la^{4\over 2-R}\\ 
&-{R\over 2-R} \mu^2\log\left(\tilde\lambda\right) +{R\over
2-R}\sum_{n=0}^\infty{\mu^{n+1}\over \tilde\la^{2(n-1)\over 2-R}} 
{\Gamma\left({n-1\over 2-R}\right)\over  
\Gamma\left({n-1\over 2-R}-n+2\right)(n+1)!}.
\end{split}
\end{equation}

{\bf (ii) Genus 1}

Proceeding with the $1/\mu$ expansion in (\ref{TODAR}) and (\ref{SCEXP})
we obtain the following torus partition function:
\begin{equation}
\label{TORUS}
\begin{split}
&F_1(\la,\mu)=\\
&{R+R^{-1}\over 24}
\left(R^{-1}X_0(y)-{2\over 2-R}\log (\tilde\la)\right)-
{1\over 24}\log \left(1-(R-1)e^{{2-R\over R} X_0(y)}\right)
\end{split}
\end{equation}

{\bf (iii) Equation for any genus $h\ge 2$}

In the same way one can obtain a linear equation relating the genus
$h$ partition function $F_h(\la,\mu)=\tilde\la^{4-4h\over 2-R}f_h(y)$
to the partition functions of all lower genera:
\begin{equation}
\label{HIGHH}
 (y\p_y+2h-2)^2 f_h-e^{-X_0(y)} \p_y^2f_h=
H( f_0,\cdots, f_{h-1}). 
\end{equation}
where $H( f_0,\cdots, f_{h-1})$ is a known function (see \cite{KKK}
for the details). The l.h.s. of this equation expressed in terms of
the variable $\xi=e^{-{1\over R}X_0(y)}$ represents a linear 2-nd
order differential operator with polynomial coefficients.

\section{ Black hole limit and  thermodynamics}

As was explained in the previous sections, the standard 2-d black hole
should correspond to $\mu=0$ and $R=3/2$. We will impose the first
condition but leave the radius $R$ arbitrary, to see the general
properties of the ``black hole'' phase $1<R<2$ and to be able to
discuss the temperature ($T={1\over 2\pi R}$) dependence of the string
partition function. For $\mu=0$ we have from the results of the
previous section the following genus expansion for the free energy
\begin{equation}
\label{BHL}
F(\lambda,\mu=0)= -{1\over 4}{(2-R)^2\over R-1}\tilde\la^{4\over 2-R}
-{R+R^{-1}\over 48}\log \left(\tilde\la^{4\over 2-R}\right)
+\sum_{h=2}^\infty\tilde\la^{4-4h\over 2-R}f_h(0)
\end{equation}
It would be very interesting to find the universal coefficients of
expansion $f_h(0)$. For that we have to solve the equations
(\ref{HIGHH}) genus by genus, using as an input the constants
$g_h(R)={\it lim}_{y\to \infty} y^{2h-2}f_h(y)$. Note that the
coefficient of the first term in the r.h.s. is also universal, as a
consequence of our approach based on the Toda equation.

For the proper black hole case we set $R=3/2$ and notice that
according to the hypothesis of \cite{FZZ} we have the following
relation between the black hole mass $M$ and the SL coupling $\la$ (in
appropriate units): $M= \tilde\lambda^8$. It means that the genus
expansion of the free energy (\ref{BHL}) takes the form of the inverse
mass expansion
\begin{equation}
\label{BHEX}
F(\lambda,\mu=0)= -{1\over 8}\ M -{13\over 288}\log M
+\sum_{h=2}^\infty M^{1-h}f_h(0)
\end{equation}

The fact that the sphere partition function given by the first term is
nonzero seems to contradict the old result obtained from the effective
dilaton gravity action for the 2-d string theory (see \cite{KATS} and
references therein). The Euclidean space-time action evaluated on the
black hole background is divergent due to linear dilaton vacuum
contribution, and to extract its universal finite part (corresponding
to the dropped non-universal terms regular in $\lambda$ in the MQM
free energy) one has to fix a subtraction procedure.  The
thermodynamic approach of \cite{GIPER} based on the leading $\alpha'$
order background (see also \cite{KATS} for its refinement based on the
exact black hole background) consists in subtracting the vacuum
contribution for fixed values of temperature $T$ and dilaton charge at
the "wall". It gives $S= M/T$ for the entropy and zero value for the
free energy F.  It was suggested in \cite{KATS} that in order to
establish the correspondence with a non-vanishing matrix model result
for F one may need an alternative reparametrization-invariant
subtraction procedure using analogy with non-critical string theory
(i.e. replacing the spatial coordinate by the dilaton field). The
subtraction of the dilaton divergence then produces a finite value for
the free energy.

It was also proposed in \cite{KATS} a microscopic estimate for the
entropy and energy of the black hole based on the contribution of
non-singlet states of the matrix model.

\section{ One- and two-point correlators from the Toda hierarchy}

The Toda integrability properties discovered in our matrix model and
summarized in Hirota bilinear relations (\ref{HIROTA}) can be used to
find the correlators of vortices of arbitrary charges defined as
\begin{equation}
\tilde K_{i_1 \cdots i_n}=\left.
\frac{\partial^{n}}{\partial t_{i_1} \cdots \partial
t_{i_n}} \log\tau_0\right|_{t_{\pm 2}=t_{\pm 3}=\cdots =0}.
\label{CORR}
\end{equation}
with $\mu$ and $t_1=-t_{-1}$ fixed.  We will give here the results for
the one- and two-point correlators in the spherical (dispersionless)
limit computed in \cite{ALKA}. Define the generating functions of the
two-point correlators
\begin{equation}
\label{TWOP}
H_{\pm}(a,b)=\sum_{m,n=0}^\infty a^nb^m\tilde Y_{\pm m,n}
\end{equation}
where we distinguish by the $\pm$ subscript the cases with vorticities
of equal or opposite sign, and of one-point correlators
\begin{equation}
\label{ONEP}
h(a)=H_\pm(a,0)=\sum_{n=0}^\infty a^n\tilde Y_{0,n}
\end{equation}
with $\tilde Y_{0,0}=0$ and
\begin{equation}
\tilde Y_{0,n}= -i\frac{1}{|n|}\frac{\partial^2 }{\partial t_n
\partial \mu} F_0  ,  \ \ \ \ 
\tilde Y_{n,m} = 2\frac{1}{|n||m|}\frac{\partial^2}
{\partial t_n \partial t_m} F_0.    
\label{Ynm}
\end{equation}
 The two-point correlators can be expressed in terms of one-point
correlators by direct analysis of Hirota equation (\ref{HIROTA}), as
was done in \cite{ALKA}, or as the correlators of free scalar field
used to formulate the Hirota identity (see the Appendix of \cite{KKK}
for the details)\footnote{We thank I. Kostov for this comment}:
\begin{equation}
\begin{split}
 H_+(a,b)&=\log\left[ \frac{4ab}{(a-b)^2}{\rm sh}^2\left(\frac{1}{2}
(h(a)-h(b)+\log\frac{a}{b})\right)\right] 
\label{PLPL} \\
 H_-(a,b)&=
2\log\left(1-Aab e^{h(a)+h(b)}\right),   
\end{split}
\end{equation}
with $A=\exp(-\p_\mu^2 F_0)$.

It is interesting that we can obtain the generating function of
one-point correlators $h(a)$ only by knowing the two point correlators
(\ref{PLPL}) and the dependence of the free energy from $t_{\pm 1}$
given by (\ref{FZERO})-(\ref{SUSC}).  Due to this we can express
$\tilde Y_{\pm 1,n}$ through $\tilde Y_{0,n}$ by  linear
integral-differential operators and write a closed equation on
$h(a)$. The solution of it is given by a simple algebraic equation
\begin{equation}
\label{SOLH}
e^{\frac{1}{R}h}-ze^{h}=1
\end{equation}
where $z=a\ \frac{\tilde\la^{R\over 2-R}}{\sqrt{R-1}}
 e^{-\frac{R-1}{R}X_0(y)}$ and $X_0(y)$ is defined by
 (\ref{SUSC}). This yields the following explicit formula for the one
 point correlators
\begin{equation}
\label{CORN}
\begin{split} 
K_n=\frac{(2-R)n\Gamma(nR+1)}{(n+1)!\Gamma(n(R-1)+2)}
\frac{\tilde\la^{\frac{nR+2}{2-R}}}{(R-1)^{n/2}}\times \\
\times\left[{(n+1)\over (2-R)n}e^{-\frac{n(R-1)+1}{R}X_0}
- {(n(R-1)+1)\over (2-R)n} e^{-(n+1)\frac{R-1}{R}X_0}\right].
\end{split}
\end{equation}
Note that $K_n\sim \la^{\frac{|n|R+2}{2-R}}\sim
\mu^{\frac{1}{2}|n|R+1}$ in perfect agreement with the scaling of
couplings $t_n\sim\mu^{-\frac{1}{2}|n|R+1}$  in (\ref{STRT}).

In the ``black hole'' limit $y\to 0$ the expression in the square
brackets equals $1$ and the formula simplifies significantly.  A
rather explicit expressions can be given also for the two point
correlators.

These correlators have to be compares (for the black hole case
$R=3/2$) to the correlators calculated in \cite{TESCH,FZZ,FAT}. Due to
the problems of identification of operators in the matrix model and
CFT formulation it is not yet done.

\section{ Conclusions }  

We presented here a matrix model describing the 2-d string theory
compactified on the circle of a radius $R$ ($R_{selfdual}<R<R_{KT}$)
with a dense gas of vortices on the world sheets. We uncovered the
Toda integrable structure of the theory and used the corresponding
Toda-Hirota equations to analyze the free energy for various genera,
as well as one- and two point correlators of vortices on the
sphere. We described the limit of the theory, corresponding (due to
the duality of \cite{FZZ}) to the two dimensional string theory on the
dilatonic black hole background.  We found that the free energy in
this limit was finite and proportional to the mass of the black hole.

There are many interesting and mysterious questions related to our
approach:

1. The Toda integrable structure found in our MQM approach was never
   discovered in the corresponding CFT (\ref{STRT}). The closest
   observation on the CFT side was the $W_{\infty}$ algebra of
   discrete states leading to the topological description of the 2-d
   string theory on the self dual radius $R=1$ \cite{MUK} giving rise
   to the Toda description \cite{EGUCHI}. Our results (and the old
   conjecture about the Toda structure in the dual version of the 2-d
   theory \cite{MPD}) suggest the existence of such an integrability
   at any $R$ in the corresponding CFT as well.

2. The supersymmetric version of the 2-d string escapes for the moment
   the similar Toda-like description, although the corresponding
   supersymmetric CFT is very similar to its bosonic counterpart.

3. The matrix approach is very promising for the clarification of 
   space time physics and of  thermodynamics of the 2-d black hole
   since we know the Hamiltonian of MQM and the origin of a large
   amount of states responsible for the great entropy of the
   underlying black hole \footnote{They are due to the ``angular''
   variables of the matrix coordinate excited in higher
   representations, as opposed to the singlet representation where
   only the eigenvalues of the matrix are relevant.}. However the
   practical calculations are still missing due to some technical
   problems. An important step on this way would be the generalization
   of the effective action of Das-Jevicki-Polchinski to the case of a
   nonzero vortex density. It would also help to see the target space
   metric of the problem and to describe the tachyonic scattering.

4. It should be also possible in our matrix approach to get a
   nonperturbative information about the black hole, trying to analyze
   the Toda equation (\ref{TODA}) with the boundary condition
   (\ref{FRNOT}) by the methods different from the genus expansion
   used until now.

\bibliographystyle{amsalpha}

\end{document}